**Title**: The determinants of COVID-19 case fatality rate (CFR) in the Italian regions and provinces: an analysis of environmental, demographic, and healthcare factors

**Author:** Gaetano Perone.
**Affiliation:** University of Bergamo
Department of Management, Economics and Quantitative Methods.
**e-mail address:** gaetano.perone@unibg.it
**ORCID**: http://orcid.org/0000-0002-0614-6727

**Abstract**. The Italian government has been one of the most responsive to COVID-2019 emergency, through the adoption of quick and increasingly stringent measures to contain the outbreak. Despite this, Italy has suffered a huge human and social cost, especially in Lombardy. The aim of this paper is dual: i) first, to investigate the reasons of the case fatality rate (CFR) differences across Italian 20 regions and 107 provinces, using a multivariate OLS regression approach; and ii) second, to build a "taxonomy" of provinces with similar mortality risk of COVID-19, by using the Ward's hierarchical agglomerative clustering method. I considered health system metrics, environmental pollution, climatic conditions, demographic variables, and three *ad hoc* indexes that represent the health system saturation. The results showed that overall health care efficiency, physician density, and average temperature helped to reduce the CFR. By the contrary, population aged 70 and above, car and firm density, level of air pollutants ($NO_2$, $O_3$, $PM_{10}$, and $PM_{2.5}$), relative average humidity, COVID-19 prevalence, and all three indexes of health system saturation were positively associated with the CFR. Population density, social vertical integration, and altitude were not statistically significant. In particular, the risk of dying increases with age, as 90 years old and above had a three-fold greater risk than the 80–to–89 years old and four-fold greater risk than 70–to–79 years old. Moreover, the cluster analysis showed that the highest mortality risk was concentrated in the north of the country, while the lowest risk was associated with southern provinces. Finally, since prevalence and health system saturation indexes played the most important role in explaining the CFR variability, a significant part of the latter may have been caused by the massive stress of the Italian health system.

**Founding**: This research did not receive any specific grant from funding agencies in the public, commercial, or not-for-profit sectors.

Highlights:

- The determinants of COVID-19 CFR in the Italian regions and provinces;
- Several environmental, demographic, and health system factors were studies;
- The methods used were OLS multivariate analysis and cluster analysis;
- $PM_{10}$, $PM_{2.5}$, $NO_2$, $O_3$, population age, humidity, and temperature directly affected the CFR;
- Saturation of the health system played an important role in explaining the CFR.

# 1. Introduction

The novel coronavirus disease (COVID-19) is a severe acute respiratory syndrome detected for the first time in December 2019 in Wuhan, Hubei province, China. On March 11, 2020, the World Health Organization (WHO, 2020) declared that COVID-19 could be characterized as a pandemic. As of July 26, 2020, according to Worldometer (2020), the virus has spread across 213 countries and territories, affecting over 16,4 million people and causing more than 650 thousand deaths. Italy was one of the countries hit the worst by the pandemic, with almost 250 thousand confirmed cases and over 35 thousand deaths at the time of writing. Despite the widely recognized excellence of the Italian health system (World Health Organization, 2010; GBD, 2017; Bloomberg 2019), the country has paid a very high price, with one of the highest case fatality rate (CFR) in the world (https://ourworldindata.org/grapher/coronavirus-cfr). However, at the peak of the epidemic, mortality was affected by a large spatial heterogeneity; in fact, the northern regions were characterized on average by a significantly higher CFR than southern regions (Figure 1).

The aim of this work is to contribute to the existing literature by investigating the main reasons and determinants of COVID-19 CFR in Italian regions and provinces. In particular, in the last months, a plenty and increasing body of literature focused its attention on the environmental, meteorological, demographic, and social factors that may affect COVID-19 mortality (Bayer and Kuhn, 2020; Brandt et al., 2020; Comunian et al., 2020; Du et al., 2020; Ma et al., 2020; Pansini and Fornacca, 2020; Sannigrahi et al., 2020; Wu et al., 2020a; Verity et al., 2020; Zhu et al., 2020).

Ma et al. (2020) used a generalized additive model (GAM) to study the relationship between the meteorological factors and the daily deaths of COVID-19 in Wuhan from January 20, 2020 to February 29, 2020. They found a positive association between the daily deaths of COVID-19 and the diurnal temperature range (DTR), and a negative relationship between the former and the relative humidity and temperature. Similarly, Wu et al. (2020b) used a log-linear GAM to analyze the effect of temperature and humidity on daily new deaths of COVID-19 in 166 countries, as of March 27, 2020. Temperature and relative humidity were found to be both significantly and negatively associated with daily new deaths. Rahman et al. (2020) performed partial correlation analysis and linear mixed effect modeling to analyze the effect of temperature on COVID-19 mortality risk in 149 countries. They showed that higher temperatures were negatively associated with mortality in high-income countries, while extreme temperature may increase mortality risk in low-and middle-income countries.

Sannigrahi et al. (2020) used spatial regression models to analyze the spatial association between the key demographic variables and COVID-19 deaths across 31 European countries. They found that the incidence of the population aged 80 and above on overall casualties caused by COVID-19 was highly significant. Verity et al. (2020) used a model-based approach to estimate the case fatality ratio associated with age groups in mainland China. They found that the population aged 80 and above had the highest case fatality ratio (13.4%). Similarly, Du et al. (2020), by implementing a univariate and multivariate logistic regression analysis, showed that advanced age is a significant risk factor for COVID-19 mortality. Moreover, Ioannidis et al. (2020), by using official data from 14 countries and 13 US states as of June 17, 2020, estimated that age risk gradients were highly significant. In particular, people aged less than 65 had 16-100-fold lower risk of COVID-19 deaths than older people.

Regarding the effect of air pollutants on COVID-19 related deaths, Conticini et al. (2020), by investigating the relevant literature, concluded that prolonged exposure to air pollutants may lead to chronic respiratory conditions, even in healthy and young people. Wu et al. (2020a) analyzed COVID-19 death counts for 3,087 counties in the USA, covering 98% of the population, by using a negative binomial mixed model. They found that a positive and significant association between $PM_{2.5}$ and COVID-19 mortality rates. In particular, a 1 unit increase in $PM_{2.5}$ is related to an 8% increase in the COVID-19 fatality rate. Pansini and Fornacca (2020), using Kendall's tau and Pearson correlation coefficient, found a positive and significant association between COVID-19 mortality and several air pollutants (CO, $NO_2$, $PM_{10}$, $PM_{2.5}$) in China and the USA.

Ogen (2020) used spatial analysis to examine the relationship between long-term exposure to $NO_2$ and COVID-19 mortality in 66 administrative regions in France, Germany, Italy, and Spain. He showed that 78% of the total COVID-19 deaths were located in north Italy and central Spain, i.e. the regions with the highest level of $NO_2$. Bianconi et al. (2020), using multiple linear regression models, showed that mean annual exposure to $PM_{10}$ and $PM_{2.5}$ was significantly and positively associated with COVID-19 death rate in Italian 20 regions. Yao et al. (2020), by using spatial analysis and multivariate linear regression for China as of April 12, 2020, found that every 10 μg/m$^3$ increase in $PM_{10}$ and $PM_{2.5}$ concentrations is associated with a 0.24% and 0.26% increase in the COVID-19 mortality rate, respectively. Brandt et al. (2020) also stressed that air pollution levels are strongly associated with densely populated urban areas in the USA. Hamidi et al. (2020) investigated both direct and indirect impacts of population density on death rates in 913 USA metropolitan countries by using structural equation modeling. They found that larger metropolitan areas and counties with higher population density were significantly associated with higher mortality rates.

In early March 2020, Bayer and Kuhn (2020) advanced the hypothesis that Italian higher vertical social integration may arise the COVID-19 fatality rate. They used a sample of 24 countries with at least 200 COVID-19 confirmed cases and found a positive correlation between the share of the population aged 30-49 living with parents and COVID-19 death rate. However, this relation was strongly criticized by Belloc et al. (2020), which showed how the sign of the correlation turned negative when considering the variation within Italian regions. Finally, several studies have shown that the presence of at least one comorbidity, such as hypertension, diabetes, cardiovascular disease, and chronic lung disease, negatively affects outcomes of patients hospitalized with COVID-19 (Guan et al., 2020; Istat-ISS, 2020; Jordan et al., 2020;Wang et al., 2020; Wu and McGoogan, 2020).

**2. Material and Method**

The aim of this study is to estimate the main determinants of COVID-19 mortality rate at the peak of the virus outbreak in 20 Italian regions and 107 Italian provinces. Specifically, I chose the peak of mild and severe cases of COVID-19, which was approximately reached on April 3–4, 2020 (Figure 2), because it can be considered as the moment of maximum health care system saturation. First, I used a multivariate cross-sectional OLS (ordinary least squares) approach to identify the main determinants at regional at province level, and then I applied the Ward's hierarchical agglomerative clustering method (Ward 1963) to build a "taxonomy" of provinces with similar mortality risk of COVID-19. The rationale behind the choice of cross-sectional regression methodology instead of a panel approach is as follows: i) comprehensive daily data on March 2020 are not currently available for environmental pollution and climatic variables; and ii) demographic and health system variables change very little or not at all in the short term.

At regional level, I used the following 17 explanatory variables (Table 1): an overall index (IPS) of the Italian health system performance in 2017–2018, the public health expenditure per capita in the period 2015–2017, the total specialist doctors and general practitioners per 1,000 inhabitants in the period 2016–2018, the total ordinary hospital beds per 1,000 inhabitants in the period 2016–2018, an index of car and firm density in 2015–2017, the electric power consumption (kWh per capita) in the period 2016–2018,[1] the proportion of population aged 70 and over in 2019, the proportion of population aged 80 and over in 2019, the proportion of population aged 90 and over in 2019, the vertical social integration proxied by the share of adults aged 18–34 living with their parents in 2019, the average relative humidity levels registered during March 2020, the average historical diurnal temperature range (DTR) in March, the average historical temperature in March, the average prevalence of COVID-19 on April 3 and 4, 2020, the ratio between the COVID-19 prevalence and the ordinary hospital beds, an two *ad hoc* indexes that indicate the saturation of ordinary hospital beds (OB) and critical care beds (CCB) at the peak of the outbreak, respectively. The dependent variable is represented by the average case fatality rate of COVID-19 on April 3 and 4, 2020.[2]

---
[1] Car and firm density and electricity consumption can be considered as two proxy indicators for air pollution.

At the province level, I used the following 18 explanatory variables (Table 2): the average general practitioners per 1,000 inhabitants in the period 2015–2017, the historical average temperature in March, the historical average diurnal temperature range (DTR) in March, an ordinal index of the population structure (rural-intermediate-urban) in 2013, the population density (people per sq. km) in 2019, the proportion of population aged 70 and over in 2019, the proportion of population aged 70–79 in 2019, the proportion of population aged 80–89 in 2019, the proportion of population aged 90 and over in 2019, the average emissions (in µg/m$^3$) of particulate matter less than 10 micrometers in diameter ($PM_{10}$) in 2017–2018, the number of days in which $PM_{10}$ exceeded the legal limit of 50 µg/m$^3$,[3] the average emissions (in µg/m$^3$) of particulate matter less than 2.5 micrometers in diameter ($PM_{2.5}$) in the period 2017–2018, the average emission of nitrogen dioxide ($NO_2$) expressed in µg/m$^3$ in 2017–2018, the number of days in which ozone ($O_3$) exceeded the limit of 120 µg/m$^3$,[4] the average altitude of the capital city of each province, the prevalence of COVID-19 on March 31, 2020, and the ratio between the COVID-19 prevalence on March 31 and the average number of ordinary hospital beds in in the period 2016–2018. The dependent variable is represented by the case fatality rate of COVID-19 on March 31, 2020.[5] Further details are provided in Tables 1 and 2.[6]

The OLS equation estimated at the regional and provincial level was as follows:

$$y_i = \beta_0 + \beta_1 X_1 + \ldots \beta_n X_n + \varepsilon_i \qquad [1]$$

Where $i$ represented the regions/provinces, $\beta_0$ was the intercept, $X_1$ to $X_n$ were the independent variables for each region/province, and $\varepsilon_i$ was the random error.

Then, I conducted a cluster analysis on selected significant variables obtained from multivariate OLS on province data. The procedure encompassed the following 4 sequential steps. First, since variables are not on the same scale, I calculated the standardized data matrix by using the following formula:

$$z = \frac{x - \mu}{\sigma} \qquad [2]$$

Where $x$ was the value of the variable in the original dataset, $\mu$ was the arithmetic mean of the original variable, and $\sigma$ was the standard deviation of the latter. Second, I computed the Euclidean distance. Given two points X and Y in d dimensional space, the Euclidean distance between X and Y was equal to:

$$\|X - Y\| = \sqrt{\sum_{i=1}^{d}(x_i - y_i)^2} \qquad [3]$$

Then, I applied Ward's hierarchical clustering method, which allowed to obtain clusters of provinces with features as similar as possible by minimizing the total within-cluster variance. Specifically, at each step, the pair of clusters that are characterized by minimum between-cluster distance were merged. Therefore, Ward's method merging cost formula between two clusters, p and q, was given by:

$$\Delta(p,q) = \sum_{i \in p \cup q} \|x_i - m_{p \cup q}\|^2 - \sum_{i \in p} \|x_i - m_p\|^2 - \sum_{i \in q} \|x_i - m_q\|^2 \qquad [4]$$

---

[2] According to Baud et al. (2020) and Scheiner et al. (2020), there is a delay between infection and death of about 14 days. However, on one hand, the prevalence on March 19–21 and April 3–4 had an almost perfectly positive correlation of 0.96 (elaboration on data from http://www.salute.gov.it), and on the other, the CFR increased exponentially until the peak, and then it has continued to grow much more slowly (https://ourworldindata.org/grapher/coronavirus-cfr).

[3] The legal limit is laid down in Directive 2008/50/EC (https://eur-lex.europa.eu/legal-content/en/ALL/?uri=CELEX%3A32008L0050).

[4] The average emissions of air pollutants for each province have been calculated by using the annual data coming from a maximum of 266 urban and suburban monitoring stations, spread all over the country (Istat, 2019).

[5] Istat released only the overall death for March 31, April 30, and May 31, 2020.

[6] Some descriptive statistics are also reported in the Supplementary Material (Tables S1 and S2).

From this, it was obtained:
$$\Delta(p,q) = \frac{n_p n_q}{n_p + n_q} \|m_p - m_q\|^2 \quad [5]$$
Where $m_j$ was the center of cluster $j$, and $n_j$ was the overall number of points included in cluster $j$.

## 4. Results and discussion
### 4.1 OLS at the regional level

In Table 3 (a, b), I presented the OLS estimations at the regional level. All the 12 estimated OLS models were statistically significant; in fact, the Fisher-Snedecor distribution assumed values far higher than the tabulated critical values at the 1% level of significance. In particular, the r-square showed that models were able to explain from 0.44% to 0.88% of CFR variability. Furthermore, since Breusch-Pagan (1979) and Shapiro-Wilk (1965) tests allowed to accept the null hypothesis of homoscedasticity and normality of residuals, models seemed well specified. However, due to the small sample, I preferred to adopt a conservative approach, by applying the HC2 correction proposed by MacKinnon and White (1985).[7] It is important to stress that OLS cross-sectional analysis is very sensitive to the presence of outliers, which can cause misspecification issues (Mur and Lauridsen 2007), especially in such small samples (Wooldridge 2015, p. 334). Therefore, I investigated the presence of highly influential points. The analysis revealed that none of the leverage points (h) are beyond the cutoff value ($h > 2n/k$) proposed by Belsley et al. (1980). Finally, I also identified the presence of multicollinearity. The variance inflation factors (VIF) ranged from 1.46 to 8.28[8] and were less than 10, i.e. the rule of thumb suggested by the relevant literature (Belsley 1982; Hair et al. 1995). Therefore, I concluded that the independent variables just suffered from weak linear dependency and there were no serious multicollinearity issues.

The results showed that the IPS index and physicians were significantly and negatively associated with COVID-19 CFR.[9] By the contrary, health expenditure, car and firm density, people aged 70 and above, humidity, DTR, prevalence, the ratio of prevalence/ordinary beds, and saturation indexes of ordinary and critical care beds were significantly and positively correlated with the CFR. Hospital beds, kWh per capita, and social vertical integration showed no statistical significance. Even if the sign of health expenditure is unexpected, the IPS index seemed to confirm the importance of health system effectiveness in all its dimensions more than only health expenditure. The regression coefficients also gave interesting information. The change in 1 unit of physician per 1,000 inhabitants and IPS index was correlated on average with a change of -2.87% and -0.53% in the CFR, respectively.[10] Consistent with Du et al. (2020), Ioannidis et al. (2020), Sannigrahi et al. (2020), the risk of dying increased progressively with the increase in the elderly population. In fact, the regression coefficients of the population aged 90 and above were 4 times larger than that for the population aged 80 and above, and 8 times larger than that for the population aged 70 and above.

Moreover, each 1 unit increase in humidity and DTR was positively associated on average with a change of 0.28%[11] and 2.22% in the CFR. The results for humidity are consistent with a recent study by Bianconi et al. (2020) on the Italian case, while DTR outcomes are in line with Ma et al. (2020). Prevalence and health system saturation indexes were highly significant in all the models. In particular, saturation of ordinary beds had the largest coefficient, followed by COVID-19 prevalence, the ratio prevalence/beds, and the saturation of intensive care beds. They played the most important role in explaining CFR variability; in fact, the average r-square of models (5–12) with prevalence and saturation indexes was 0.81, i.e. almost twice the models (1–4) without them (r = 0.46). In particular, the change in 0.1 unit of CCB and OB saturation was correlated with a change of 0.47% and 4.54% in the CFR. Therefore, a significant part of the CFR may be caused by the massive stress of the Italian health system.

---

[7] As suggested by Erving and Long (2000, p. 220), in presence of homoskedasticity, HC2 option has good small size properties.

[8] I did not consider VIF for average temperature, which was constantly greater than 10. It is due to this that average temperature has been excluded in most part of the models.

[9] In addition, in four OLS models, the average electric power consumption is negatively associated with the CFR. However, its coefficients are statistically significant only at the 10% level.

[10] From now on, I only considered the average of the significant coefficients.

[11] Model 10 also seemed to suggest that the relationship between humidity and CFR is non-monotonic. In other words, the CFR decreased for low-medium levels of humidity and increased for high values of humidity. However, since the lack of comprehensive data on humidity, further investigations are necessary.

Finally, the correlation matrix in Figure S1 indicated that the average temperature and humidity were significantly and negatively correlated to the COVID-19 prevalence, with a Pearson's r of -0.8 and -0.46, respectively. By the contrary, the kWh per capita and car and firm density showed a good positive correlation with the latter, with a Pearson's r of 0.66 and 0.52, respectively.

## 4.2 OLS at the provincial level

In Table 4 (a, b), I presented the OLS estimations at the provincial level. All 11 estimated OLS models were statistically significant; in fact, the Fisher-Snedecor distribution assumed values far higher than the tabulated critical values at 1% level of significance. In particular, r-square showed that models were able to explain from 0.27% to 0.45% of CFR variability.

Since the Breusch-Pagan (1979) test revealed heteroscedasticity issues, I applied the HC2 correction. As stated by Ghasemi and Zahediasl (2012), the violation of normality assumption should not be a major problem in a sample with enough observations (n > 40), such as in this case. Furthermore, the variance inflation factors (VIF) ranged from 1.05 to 2.56 and are much less than 10; thus, there were no multicollinearity issues.

The OLS models showed that general practitioners, average temperature, and DTR were significantly and negatively associated with the CFR.[12] By the contrary, all the considered air pollutants ($PM_{10}$, $PM_{2.5}$, $NO_2$, and $O_3$), people aged 70 and above, prevalence, and saturation index for ordinary beds were significantly and positively correlated with the CFR. Urbanization, population density, and altitude showed no statistical significance. The change in 1 degree Celsius of temperature and DTR was correlated with an average change of -0.53% and -2.44% in the CFR, respectively. If the results for the temperature are consistent with other studies (Bianconi et al. 2020; Rahman et al. 2020; Wu et al. 2020b), the sign of DTR is unexpected and in contrast with previous findings. Therefore, the relative impact is ambiguous and unclear.

Among air pollutants, $PM_{2.5}$ and $PM_{10}$ had the largest impact on COVID-19 CFR, followed by $NO_2$ and $O_3$. The change in 10 μg/m$^3$ of $PM_{2.5}$ and $PM_{10}$ was associated with a change of 3.13% and 2.54% in the CFR, respectively. The change in 10 μg/m$^3$ of $NO_2$ was "only" correlated with a change of 1.27% in the CFR. While, each day in which $O_3$ exceeded the limit of 120 μg/m$^3$, the CFR have increased of 0.11%. The inclusion of $O_3$ is very significant, since it allowed r-square to increase to 0.42. The positive and significant relationship between air pollutants and CFR is consistent with the relevant literature (Bianconi et al., 2020; Ogen, 2020; Pansini and Fornacca, 2020; Wu et al., 2020a; Yao et al., 2020). As found at the regional level, the CFR increased progressively with population age. Specifically, 90 years old and above had a three-fold greater risk than the 80–to–89 years old and four-fold greater risk than 70–to–79 years old. This result is consistent with Onder et al. (2020) and assumes great importance, since the susceptibility to the COVID-19 is constant across all age groups (Istat-ISS, 2020). The prevalence and saturation indexes were highly and significantly correlated with the CFR in all the considered models (9–12). Every 0.1 unit increase in COVID-19 prevalence and saturation of hospital ordinary beds was associated with an average change of 1.66% and 0.44% in the CFR, respectively. Moreover, the inclusion of prevalence and saturation in the models allowed an increase in the average r-square from 0.31 to 0.43. This is consistent with the previous findings.

The correlation matrix (Figure S2) gave other interesting information, especially regarding air pollutants. Air pollutants were highly and positively correlated with each other, with a Pearson's r that ranged from 0.48 to 0.84. Moreover, consistent with Brandt (2020), population density was moderately and positively correlated with air pollutants. Finally, as found in other studies (Bianconi et al. 2020; Pansini and Fornacca 2020; Setti et al. 2020; Zhu et al. 2020), COVID-19 prevalence and air pollutant showed a good positive correlation, with a Pearson's r ranged from 0.36 to 0.65.

---

[12] It is important to stress that signs for temperature and DTR didn't change even considering the historical values of winter. This was also true for the regional analysis.

**4.3 Cluster Analysis**

In Figure 3, I presented the dendrogram obtained using Ward's method. The optimal number of clusters was identified by using two different methods: i) the EM (Expectation-Maximization) algorithm for Gaussian finite mixture model, proposed by Fraley et al. (2012) in the package 'Mclust' (R environment); and iii) the package 'NbClust' (R environment) that proposes 30 different indexes.[13] In the first case, according to BIC (Bayesian Information Criterion) score, the best model was VVI (varying volume, varying shape, and equal orientation) with 3 clusters. The majority of indices (7 out of 30) included in 'NbClust', also proposed 3 clusters. Therefore, by cutting the dendrogram at an approximately height of 13, I obtained three clusters with an increasing risk of mortality (Table 5). Each cluster was identified with a grey dotted rectangle (Figure 3) and represented graphically through a map (Figure 4). The cluster with the highest mortality risk (risk = 3) was composed by 25 provinces of northern Italy, 1 province of central Italy (Pesaro and Urbino), and 2 provinces of southern Italy (Frosinone and Avellino). The cluster with the medium mortality risk (risk = 2) was composed by 22 provinces of northern Italy, 18 provinces of central Italy, and 6 provinces of southern Italy (Chieti, L'Aquila, Oristano, Pescara, South Sardinia, and Teramo). Finally, the cluster with the lowest mortality risk (risk = 1) was composed by 2 provinces of central Italy (Latina and Rome), and 31 provinces of southern Italy (Table 6). Therefore, the highest mortality risk was concentrated in the north of the country, while the lowest risk was associated with southern provinces. Specifically, cluster 3 had an 8.62% higher CFR than cluster 1, and 5.3% higher CFR than cluster 2. Cluster 3 had a temperature of 2.58 degrees lower and a number of general practitioners per 1,000 inhabitants of 0.24 lower than that for cluster 1. Moreover, cluster 3 had a $PM_{10}$ of 11.23 μg/m$^3$ higher and a proportion of population aged 70 and abobe of 1.49% higher than that for cluster 1. Finally, the hospital bed saturation in cluster 3 was more than double than that for cluster 2, and 8 times greater than that for cluster 1. The results are consistent with Grasselli et al. (2020), according to which Lombardy's intensive care units already had an 85% to 90% occupancy ahead of the outbreak.[14]

**5. Conclusions**

To the best of my knowledge, this is one of the first studies to investigate the relationship between a wide set of heterogeneous factors and COVID-19 mortality. The OLS analysis showed that environmental, demographic, and healthcare factors play an important role in explaining the CFR variability. In particular, population aging, air pollutants ($NO_2$, $O_3$, $PM_{10}$, and $PM_{2.5}$), relative humidity, COVID-19 prevalence, and critical care and ordinary beds saturation are positively correlated with the CFR. By the contrary, overall health care efficiency (IPS), physician density, and average temperature are negatively associated with CFR. Specifically, the inclusion of the COVID-19 prevalence and saturation indexes of ordinary and critical care beds explains up to 86% of the CFR variability. Therefore, a significant part of the CFR variability may be caused by the massive stress of the Italian health system. The results are robust across several model specifications. Moreover, cluster analysis showed that the highest mortality risk was concentrated in northern Italy, while the lowest risk was associated with southern provinces.

However, this study also has some limitations that can be summarized as follows: i) first, a significant part of the patients died in hospital presented at least one comorbidity ahead of COVID-19 infection (ISS, 2020); ii) then, as pointed out in other studies (Spalt et al., 2016; Wang et al., 2019), the utilization of air pollution implies unavoidable measurement errors since most people usually stay indoors; and iii) finally, climatic variables, such as average temperature and DTR, refer just to the average historical values.

---

[13] The outcomes of both methods are reported in the Supplementary Material (Figure S3 and Table S3).
[14] In fact, Lombardy was the hardest hit region on April 3–4, 2020, with 40% of the Italian confirmed cases.

Finally, the study seemed to stress the importance of implementing quick and rational lockdown measures, of making patients comfortable, of implementing an action plan to discourage car use and decrease firm's pollution, and of buying *ad hoc* health care facilities, medical equipment, and devices to adequately tackle similar and unforeseeable emergencies.

Table 1. Definitions of all variables used for OLS regional analysis.

| Variables | Definitions | Sources |
|---|---|---|
| *Dependent variables* | | |
| CFR | The average case fatality rate for COVID-19 in each region, obtained by dividing the average confirmed deaths by the average confirmed cases on April 3 and 4, 2020. | Italian Ministry of Health[15] |
| *Independent variables* | | |
| IPS | A synthetic index of the Italian health system performance in the period 2017–2018, which includes eight different parameters.[16] | Demoskopika Research Institute (2018, 2019) |
| Health expenditure | The average public health expenditure per capita for each region, in the period 2015–2017. | I.Stat (database)[17] |
| Physicians | The average total specialist doctors and general practitioners (per 1,000 inhabitants) for each region, in the period 2016–2018. | I.Stat (database) |
| Hospital beds | The average ordinary hospital beds (per 1,000 inhabitants) for each region, in 2016–2018. | Italian Ministry of Health |
| Cars & Firms | A synthetic index of car and firm (> 250 employees) density for each region, in 2015–2017.[18] | I.Stat (database) |
| kWh per capita | The average electric power consumption in kilowatt-hours (kWh) per capita for each region, in the period 2016–2018. | Terna (2019) |
| Ages 70+ | The proportion of population aged 70 and over for each region, in 2019. | I.Stat (database) |
| Ages 80+ | The proportion of population aged 80 and over for each region, in 2019. | I.Stat (database) |

---

[15] Data are available at URL: www.salute.gov.it.
[16] The parameters used are the following: patient satisfaction, active patient mobility, passive patient mobility, legal fees for disputes, operating result, life expectancy, equality in health treatment, and economic hardship. In particular, each parameter is standardized, with mean = 100 and standard deviation = 10, and the final synthetic index is obtained by calculating the simple average of them.
[17] Data are available at URL: http://dati.istat.it/.
[18] The number of cars refers to those recorded in the Pubblico registro automobilistico (Public vehicle register). The number of largest firms (> 250 employees) refers to those that operate in the following sectors: (1) mining and minerals from quarries and mines; (2) manufacturing activities; (3) supply of electricity, gas, vapors, and air conditioning; and (4) supply sewerage, waste management and remediation activities. The index is compiled according to the following analytical method: i) first, I standardized the data according to surface area (cars and firms for 100 sq. km.); ii) than, the respective outputs are switched to fixed-base indexes (with mean = 100); iii) finally, I computed the simple arithmetic mean of the latter.

| Ages 90+ | The proportion of population aged 90 and over for each region, in 2019. | I.Stat (database) |
| Vertical integration | The share of unmarried young adults aged 18–34 living with at least one parent for each region, in 2019. | I.Stat (database) |
| Humidity | The average relative humidity levels registered during March 2020, for each region.[19] | www.il meteo.it[20] |
| DTR | The historical diurnal temperature range in March, for each region. | Mipaaf (2019a) |
| Temperature | The historical average temperature in March, for each region. | Mipaaf (2019a) |
| Prevalence | The average ratio between the people who have been tested positive for COVID-19 and the overall population of each region on April 3 and 4, 2020. | I.Stat (database), Ministry of Health (2020) |
| Preval./Beds | The ratio between the average COVID-19 prevalence on April 3 and 4, 2020, and the average number of ordinary hospital beds in 2016–2018, for each region. | Ministry of Health (2020) |
| CCB saturation | The ratio between the average people who have been recovered from COVID-19 in intensive care on 3 and 4 April 2020, and the average number of critical care beds (CCB) in the period 2016–2018, for each region. | Ministry of Health (2020) |
| OB saturation | The ratio between the average people who have been recovered from COVID-19 with mild symptoms on 3 and 4 April 2020, and the average number of ordinary hospital beds in the period 2016–2018 for each region. | Ministry of Health (2020) |

Table 2. Definitions of all variables used for OLS province analysis.

| **Variables** | **Definitions** | **Sources** |
|---|---|---|
| *Dependent variables* | | |
| Death rate | The average case fatality rate for COVID-19 in each province, obtained by dividing the confirmed deaths by the number of confirmed cases, on 31 March 2020. | Istat-ISS (2020) |
| *Independent variables* | | |
| General practitioners | The average general practitioners for each province, in 2019. | Il Sole 24 Ore (2019) |
| Temperature | The historical average temperature in March, for each province. | Mipaaf (2019b) |
| DTR | The historical average diurnal temperature range in March, for each province. | Mipaaf (2019b) |
| Urbanization | | |

---

[19] The average values have been calculated by dividing the data coming from 62 different official weather stations, managed by the Italian Air force and located in the main Italian provinces.

[20] This is one of the most trusted Italian weather forecast website. https://www.ilmeteo.it/business/assets//images/aboutUs/pdf/Google%2009_04_2020.pdf.

| | | |
|---|---|---|
| Density | An ordinal index that ranks population of each province by urban-rural structure: predominantly rural (1), intermediate (2), and predominantly urban (3). | European Commission (2013) |
| | The number of human inhabitants per square kilometer (sq. km.) of land area for each province, in 2019. | I.Stat (database) |
| Ages 70+ | The proportion of population aged 70 and over for each province, in 2019. | I.Stat (database) |
| Ages 70–79 | The proportion of population aged 70–79 for each province, in 2019. | I.Stat (database) |
| Ages 80–89 | The proportion of population aged 80–89 for each province, in 2019. | I.Stat (database) |
| Ages 90+ | The proportion of population aged 90 and over for each province, in 2019. | I.Stat (database) |
| $PM_{10}$ | The average emissions of particulate matter less than 10 micrometers in diameter, expressed in µg/m$^3$ for each province, in the period 2017–2018. | Istat (2019) |
| $PM_{10}$ (>50) | The average number of days in which $PM_{10}$ exceeded the limit of 50 µg/m$^3$ for each province, in the period 2017–2018. | Istat (2019) |
| $PM_{2.5}$ | The average emissions of particulate matter less than 2.5 micrometers in diameter, expressed in µg/m$^3$ for each province, in the period 2017–2018. | Istat (2019) |
| $NO_2$ | The average emissions of nitrogen dioxide, expressed in µg/m$^3$ for each province, in the period 2017–2018. | Istat (2019) |
| $O_3$ | The average number of days in which ozone exceeded the limit of 120 µg/m$^3$ for each province, in the period 2017–2018. | Istat (2019) |
| Altitude | The average altitude of the capital city of each province. | Istat (2019) |
| Prevalence | The ratio between the people who have been tested positive for COVID-19 on March 31, 2020, and the total population of each province in 2019 | I.Stat (database), Italian Ministry of Health |
| OB saturation | The ratio between the COVID-19 prevalence on March 31, 2020, and the average number of ordinary hospital beds in 2016–2018, for each province. | Italian Ministry of Health |

Table 3a (models 1–6). OLS regression at the regional level between CFR and environmental, demographic, and healthcare factors.

| Variables | Model 1 | Model 2 | Model 3 | Model 4 | Model 5 | Model 6 |
|---|---|---|---|---|---|---|
| Constant | 16.4788 | 21.4019 | 27.8448* | -12.1666 | -11.7866 | -14.8309 |
|  | [13.9432] | [12.8697] | [12.7974] | [28.5844] | [12.8946] | [13.3257] |
| IPS | -0.4788* | -0.5451** | -0.5731** | -0.5758** | -0.4511** | -0.4808*** |
|  | [0.2303] | [0.2221] | [0.2153] | [0.2034] | [0.1402] | [0.1468] |
| Health exp. | 0.0129** | 0.0157** | 0.0144** | 0.0237** | 0.0146* | 0.0168** |
|  | [0.0059] | [0.0056] | [0.0058] | [0.0096] | [0.0074] | [0.0067] |
| Physicians | -3.2988** | -3.1879** | -3.1683** | -2.3358 | -3.0878** | -2.5227* |
|  | [1.2894] | [1.1264] | [1.136] | [1.8958] | [1.3113] | [1.1406] |
| H. Beds | 3.8867 | 2.7984 | 3.2841 | -1.0369 | -1.4253 | -2.1594 |
|  | [2.8595] | [2.8415] | [3.1279] | [5.1024] | [1.7597] | [2.191] |
| Car & Firm | 6.2449*** | 7.0056*** | 6.9565*** | 8.5611*** | 4.5281** | 5.4669*** |
|  | [1.7936] | [1.7842] | [1.93] | [1.9709] | [1.4903] | [1.4142] |
| Kilowatt | -0.0007 | -0.0005 | -0.0005 | -0.0015* | -0.0009 | -0.0012* |
|  | [0.0006] | [0.0006] | [0.0007] | [0.0007] | [0.0007] | [0.0006] |
| Aged 70+ | 0.7821** |  |  | 1.0916** | 0.7973*** | 0.8982*** |
|  | [0.3183] |  |  | [0.4199] | [0.2343] | [0.234] |
| Aged 80+ |  | 1.5** |  |  |  |  |
|  |  | [0.5131] |  |  |  |  |
| Aged 90+ |  |  | 6.3807** |  |  |  |
|  |  |  | [2.5141] |  |  |  |
| Humidity |  |  |  | 0.449** | 0.1413 | 0.2615** |
|  |  |  |  | [0.1886] | [0.1535] | [0.0808] |
| DTR |  |  |  | 0.3657 | 2.679*** | 2.3164*** |
|  |  |  |  | [1.5456] | [0.6612] | [0.6588] |
| Temperature |  |  |  | -1.1144* | 0.5302 |  |
|  |  |  |  | [0.5752] | [0.4919] |  |
| Prevalence |  |  |  |  | 26.5164*** | 22.0411*** |
|  |  |  |  |  | [4.3121] | [5.3084] |
| Breusch-P. (p) | 0.2454 | 0.5101 | 0.3985 | 0.7505 | 0.4038 | 0.6391 |
| Shapiro-W. (p | 0.9892 | 0.9969 | 0.9713 | 0.7997 | 0.0115 | 0.1289 |
| Influential (h) | 0.13–0.78 | 0.15–0.74 | 0.17–0.69 | 0.23–0.82 | 0.25–0.87 | 0.2–0.87 |
| VIF | 1.56–4.44 | 1.46–4.45 | 1.49–4.88 | 2.49–5.43 | 2.65–8.28 | 2.04–5.79 |
| F-statistic | 5.85*** | 5.28*** | 5.5*** | 5.18*** | 22.13*** | 22.55*** |
| Observations | 20 | 20 | 20 | 20 | 20 | 20 |
| Adjusted $R^2$ | 0.4425 | 0.4753 | 0.4531 | 0.4899 | 0.8102 | 0.8127 |

Notes: h, leverage; p, p-value. Standard errors (in brackets) are based on HC2 method developed by MacKinnon and White (1985). Significance level: p-value < 0.01***; p-value < 0.05**: p-value < 0.1*.

Table 3b (models 7–12). OLS regression at the regional level between CFR and environmental, demographic, and healthcare factors.

| Variables | Model 7 | Model 8 | Model 9 | Model 10 | Model 11 | Model 12 |
|---|---|---|---|---|---|---|
| Constant | -9.9732 | -3.5934 | -12.2744 | 168.707** | -4.8115 | 8.2251 |
|  | [12.8992] | [12.9892] | [13.7815] | [65.3856] | [20.9305] | [12.6372] |
| IPS | -0.5065*** | -0.5475*** | -0.4995*** | -0.671*** | -0.6399*** | -0.4174** |
|  | [0.1497] | [0.1504] | [0.1506] | [0.1663] | [0.1731] | [0.1348] |
| Health exp. | 0.0151** | 0.0155** | 0.0163** | 0.012** | 0.0156 | 0.0062 |
|  | [0.0056] | [0.0056] | [0.0069] | [0.0051] | [0.0099] | [0.0074] |
| Physicians | -2.4044* | -1.8737 | -2.4275* | -1.8232 | -1.838 | -0.8314 |
|  | [1.2974] | [1.2409] | [1.1621] | [1.0917] | [1.3513] | [0.9748] |
| H. Beds |  |  | -0.7423 | -1.2938 | 1.3549 | -1.0043 |
|  |  |  | [2.3357] | [1.4252] | [3.2707] | [2.4216] |
| Car & Firms | 5.2252*** | 5.7172*** | 5.395*** | 7.2545*** | 6.4978*** | 0.6217 |
|  | [1.391] | [1.4174] | [1.4313] | [1.2172] | [1.4789] | [1.6553] |
| Kilowatt | -0.0012 | -0.0009 | -0.0012* | -0.001* | -0.0007 | 0.0000 |
|  | [0.0007] | [0.0007] | [0.0006] | [0.0004] | [0.0006] | [0.0004] |
| Aged 70+ | 0.7931** |  | 0.7943** |  | 0.8673** | 0.2744 |
|  | [0.263] |  | [0.2501] |  | [0.3488] | [0.3046] |
| Aged 80+ |  | 1.3558** |  | 2.1616*** |  |  |
|  |  | [0.4649] |  | [0.6187] |  |  |
| Vertical Integ. | -0.008 | 0.0254 |  |  |  |  |
|  | [0.1366] | [0.1343] |  |  |  |  |
| Humidity | 0.2455** | 0.2108** | 0.2536** | -4.6525** | 0.2824** | -0.0854 |
|  | [0.0956] | [0.0927] | [0.084] | [1.8055] | [0.1213] | [0.0711] |
| Humidity^2 |  |  |  | 0.036** |  |  |
|  |  |  |  | [0.0134] |  |  |
| DTR | 1.9813** | 1.6247** | 2.0091** | 2.5458*** | 1.18 | 2.3931** |
|  | [0.7199] | [0.6522] | [0.6946] | [0.6437] | [1.0982] | [0.8386] |
| Preval./beds | 7.2081*** | 7.121*** | 7.3835*** | 6.9567*** |  |  |
|  | [2.0728] | [2.0874] | [1.8403] | [1.1719] |  |  |
| ICB saturation |  |  |  |  | 4.6762** |  |
|  |  |  |  |  | [1.5899] |  |
| OB saturation |  |  |  |  |  | 45.4811*** |
|  |  |  |  |  |  | [7.9362] |
| Breusch-P. (p) | 0.5386 | 0.6034 | 0.6502 | 0.5497 | 0.8678 | 0.645 |
| Shapiro-W (p) | 0.087 | 0.8456 | 0.0951 | 0.9033 | 0.1169 | 0.6978 |
| Influential (h) | 0.2–0.85 | 0.21–0.85 | 0.21–0.87 | 0.23–-92 | 0.2–0.81 | 0.2–0.81 |
| VIF | 2.32–5.89 | 2.08–5.91 | 1.95–5.43 | - | 2.31–5.48 | 2.03–7.25 |
| F-statistic | 18.96*** | 25.07*** | 21.32*** | 20.24*** | 9.57*** | 54.14*** |
| Observations | 20 | 20 | 20 | 20 | 20 | 20 |
| Adjusted $R^2$ | 0.7965 | 0.799 | 0.7982 | 0.8782 | 0.7024 | 0.8576 |

Notes: h, leverage; p, p-value. Standard errors (in brackets) are based on HC2 method developed by MacKinnon and White (1985). Significance level: p-value < 0.01***; p-value < 0.05**: p-value < 0.1*.

Table 4a (models 1–6). OLS regression at the provincial level between CFR and environmental, demographic, and healthcare factors.

| Variables | Model 1 | Model 2 | Model 3 | Model 4 | Model 5 | Model 6 |
|---|---|---|---|---|---|---|
| Constant | 21.3371* | 22.645** | 21.4665** | 28.7773*** | 27.8601** | 28.3833** |
|  | [10.8017] | [11.0182] | [10.4233] | [10.0096] | [11.075] | [11.7957] |
| G. P. | -10.0768** | -9.4475** | -10.7185** | -10.8571** | -9.9708** | -12.8379*** |
|  | [4.2187] | [4.1747] | [4.298] | [4.2279] | [4.0876] | [4.2603] |
| Temperature | -0.5506** | -0.5966** | -0.5292** | -0.5296** | -0.4515* | -0.4038 |
|  | [0.242] | [0.236] | [0.2431] | [0.2449] | [0.2455] | [0.255] |
| DTR | -2.5289** | -2.6147** | -2.3998** | -2.6982** | -2.516** | -2.5241** |
|  | [1.0323] | [1.0539] | [1.0138] | [1.0269] | [1.0833] | [1.049] |
| $PM_{10}$ (µg/m$^3$) | 0.2569*** | 0.2529*** | 0.256*** | 0.252*** |  |  |
|  | [0.0939] | [0.0952] | [0.0938] | [0.0931] |  |  |
| Urbanization | 0.8979 | 0.7571 | 1.0869 | 1.0815 |  |  |
|  | [0.9731] | [0.9805] | [0.97] | [0.9602] |  |  |
| Pop. Density | -0.0004 | -0.0005 | -0.0004 | -0.0003 |  |  |
|  | [0.0012] | [0.0012] | [0.0012] | [0.0012] |  |  |
| Aged 70+ | 0.7596** |  |  |  | 0.6823** | 0.6955** |
|  | [0.3106] |  |  |  | [0.3157] | [0.3149] |
| Aged 70-79 |  | 1.2866** |  |  |  |  |
|  |  | [0.6163] |  |  |  |  |
| Aged 80-89 |  |  | 1.9984*** |  |  |  |
|  |  |  | [0.7511] |  |  |  |
| Aged 90+ |  |  |  | 5.5753** |  |  |
|  |  |  |  | [2.2503] |  |  |
| $PM_{10}$ ( > 50) |  |  |  |  | 0.0739*** |  |
|  |  |  |  |  | [0.0257] |  |
| $NO_2$ (µg/m$^3$) |  |  |  |  |  | 0.1268** |
|  |  |  |  |  |  | [0.055] |
| Breusch-P p. | 0.0334 | 0.05 | 0.043 | 0.0385 | 0.0049 | 0.0083 |
| F-statistic | 6.37*** | 5.9*** | 6.33*** | 6.95*** | 9.81*** | 9.0808*** |
| VIF | 1.11–1.56 | 1.09–1.56 | 1.12–1.55 | 1.18–1.57 | 1.06–1.43 | 1.07–1.52 |
| Observations | 107 | 107 | 107 | 107 | 107 | 107 |
| Adjusted $R^2$ | 0.2944 | 0.2775 | 0.306 | 0.2903 | 0.301 | 0.266 |

Notes: p, p-value. Standard errors (in brackets) are based on HC2 method developed by MacKinnon and White (1985). Significance level: p-value < 0.01***; p-value < 0.05**: p-value < 0.1*.

Table 4b (models 7–12). OLS regression at the provincial level between CFR and environmental, demographic, and healthcare factors.

| Variables | Model 7 | Model 8 | Model 9 | Model 10 | Model 11 | Model 12 |
|---|---|---|---|---|---|---|
| Constant | 31.4907*** | 41.079*** | 14.7429 | 18.1456 | 15.5861 | 18.8624 |
|  | [9.6085] | [11.7581] | [11.7796] | [11.3803] | [11.579] | [11.2617] |
| G. P. | -12.4582*** | -13.0142*** | -7.0346** | -7.6595** | -6.9201* | -7.578** |
|  | [4.642] | [4.2388] | [3.4178] | [3.5788] | [3.5378] | [3.7071] |
| Temperature | -0.22 | -0.4403* | -0.1162 | -0.129 | -0.2588 | -0.2667 |
|  | [0.262] | [0.2471] | [0.2952] | [0.2905] | [0.2778] | [0.2745] |
| DTR | -2.7569*** | -4.0316*** | -1.7712* | -1.7872* | -1.8775* | -1.8874* |
|  | [1.0324] | [1.2007] | [1.0379] | [1.0266] | [1.055] | [1.0434] |
| $PM_{10}$ ($\mu g/m^3$) |  |  | 0.0476 | 0.0454 | 0.0961 | 0.0939 |
|  |  |  | [0.0476] | [0.09] | [0.0989] | [0.0989] |
| Aged 70+ | 0.5006 | 0.6215* | 0.64** |  | 0.6583** |  |
|  | [0.3969] | [0.3324] | [0.3028] |  | [0.3031] |  |
| Aged 80+ |  |  |  | 1.178** |  | 1.2287** |
|  |  |  |  | [0.5477] |  | [0.5514] |
| $O_3$ ( > 120) | 0.1084*** |  |  |  |  |  |
|  | [0.0293] |  |  |  |  |  |
| $PM_{2.5}$ ($\mu g/m^3$) |  | 0.3127*** |  |  |  |  |
|  |  | [0.1112] |  |  |  |  |
| Altitude |  |  | -0.0002 | -0.0005 | -0.0002 | -0.0005 |
|  |  |  | [0.0021] | [0.0021] | [0.0022] | [0.0022] |
| Prevalence |  |  | 16.7325*** | 16.5201*** |  |  |
|  |  |  | [3.7856] | [3.842] |  |  |
| OB saturation |  |  |  |  | 4.4258*** | 4.3711*** |
|  |  |  |  |  | [1.0094] | [1.0332] |
| Breusch-P p. | 0.0003 | 0.0045 | 0.0000 | 0.0000 | 0.0003 | 0.0004 |
| F-statistic | 14.22*** | 10.48*** | 11.67*** | 11.77*** | 14.24*** | 14.14*** |
| VIF | 1.07–1.54 | 1.05–1.33 | 1.13–2.52 | 1.14–2.5 | 1.13–2.44 | 1.14–2.42 |
| Observations | 88 | 92 | 107 | 107 | 107 | 107 |
| Adjusted $R^2$ | 0.4241 | 0.3531 | 0.4488 | 0.4492 | 0.4142 | 0.416 |

Notes: p, p-value. Standard errors (in brackets) are based on HC2 method developed by MacKinnon and White (1985). Significance level: p-value < 0.01***; p-value < 0.05**: p-value < 0.1*.

Table 5. The clusters obtained by cutting dendrogram at an approximately height of 13.

| Variables | Cluster (CL1) (Low risk) | Cluster (CL2) (Medium risk) | Cluster (CL3) (High risk) | CL3 - CL1 |
|---|---|---|---|---|
| CFR | 4.5758 | 7.8961 | 13.1957 | 8.6199 |
| G. practitioners | 1.0322 | 0.939 | 0.7947 | -0.2375 |
| Temperature | 10.2758 | 7.1516 | 7.6946 | -2.5812 |
| Aged 70+ | 15.9355 | 19.3297 | 17.4285 | 1.493 |
| $PM_{10}$ ($\mu g/m^3$) | 21.9211 | 21.9833 | 33.1464 | 11.2253 |
| OB saturation | 0.1393 | 0.5019 | 1.161 | 1.0217 |
| Numerosity | 33 | 46 | 28 | - |

Figure 1. The case fatality rate (CFR) at the peak of the epidemic in the Italian provinces.

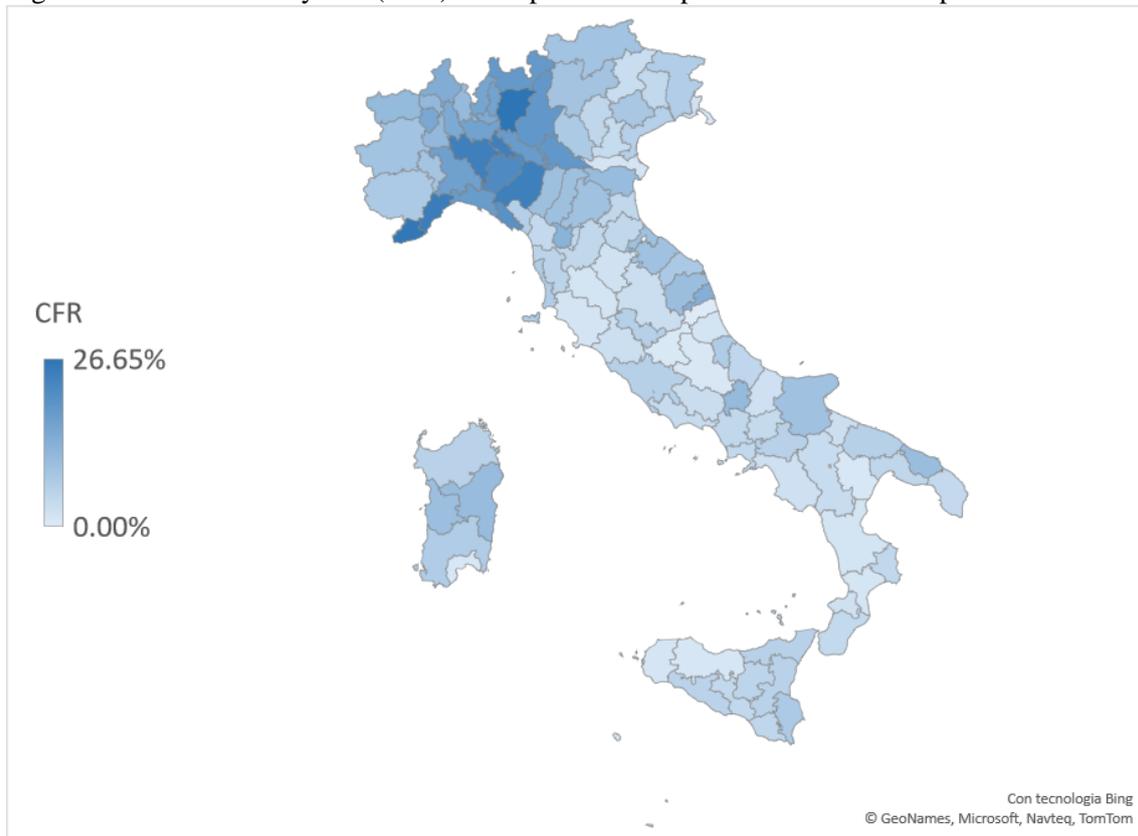

Source: Italian Ministry of Health (www.salute.gov.it).

Figure 2. The number of people recovered from COVID-19 in the period February 23, 2020–July 23, 2020.

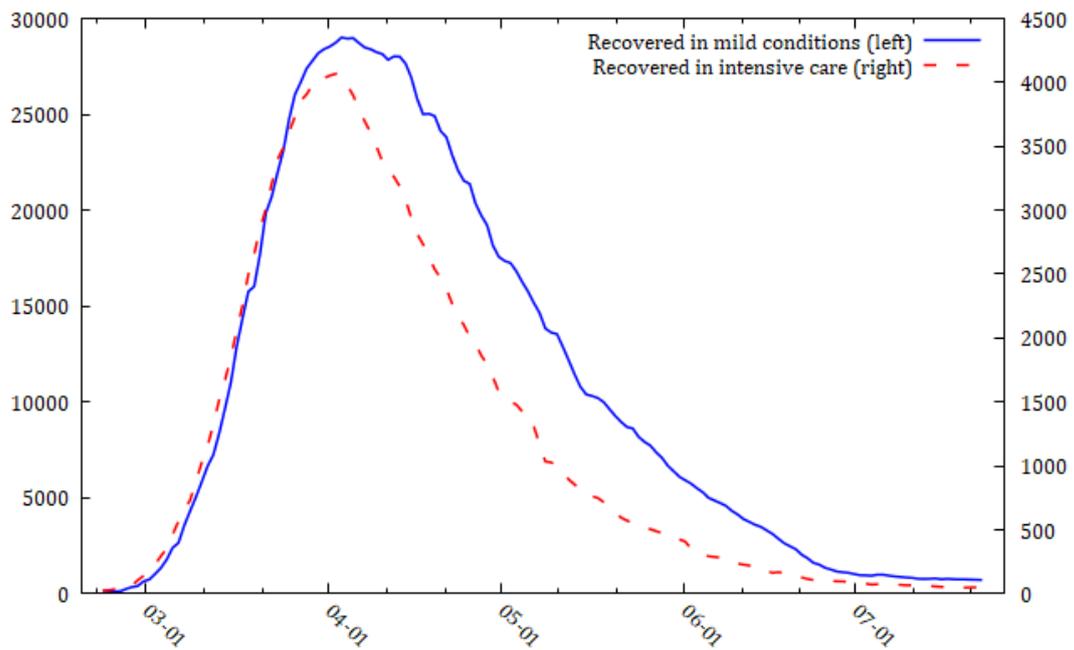

Source: Italian Ministry of Health (www.salute.gov.it).

Figure 3. Dendrogram of provinces obtained using Ward's method.

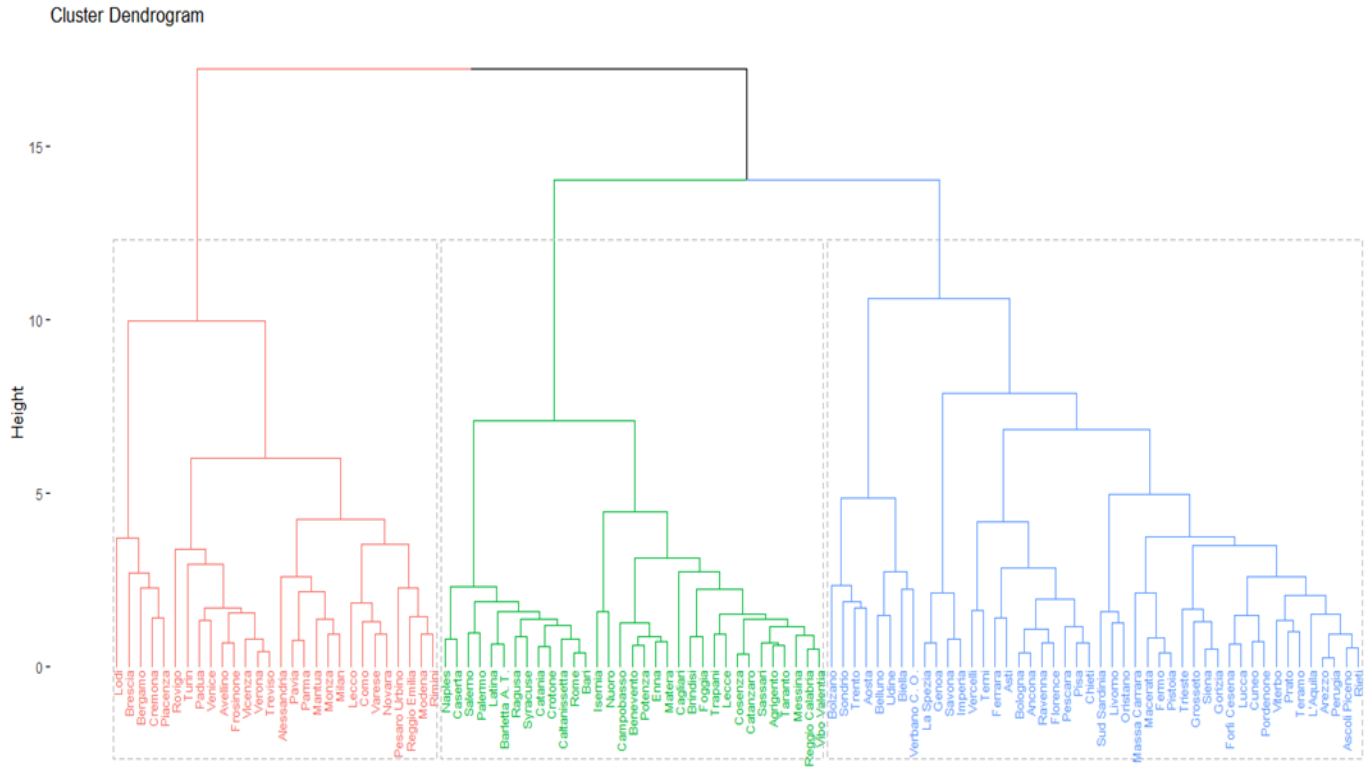

Figure 4. Map of the 107 Italian provinces divided into three increasing clusters of risk.

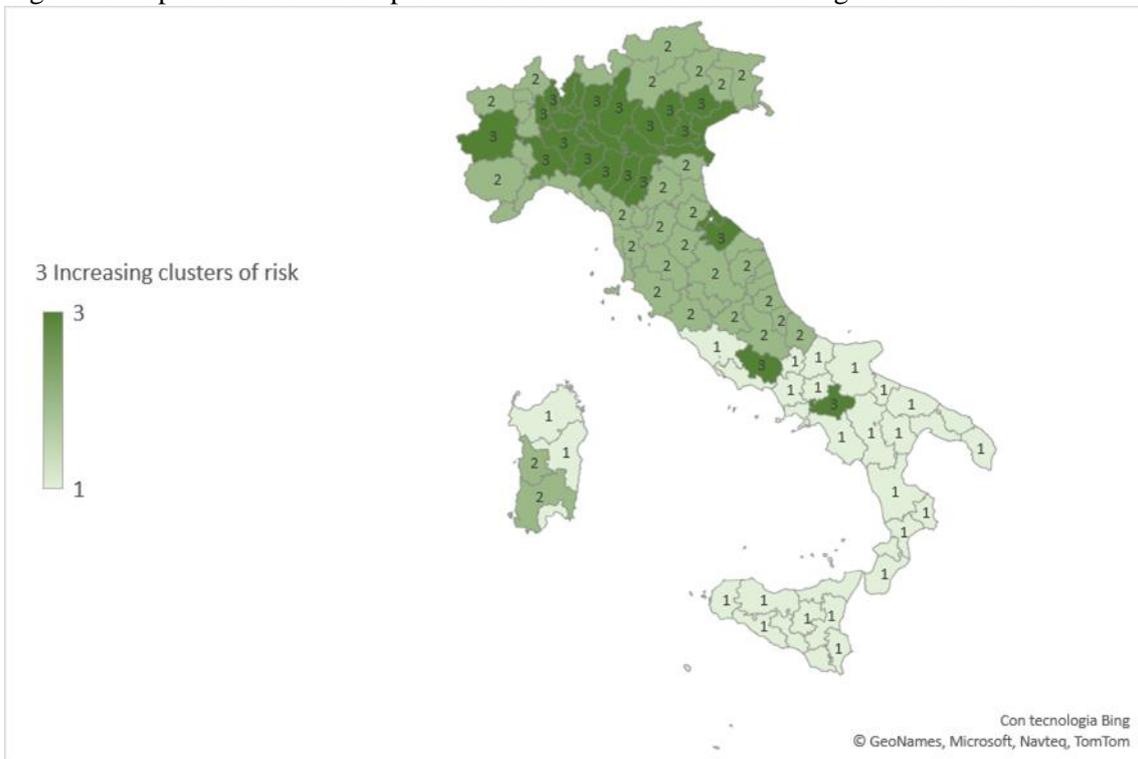

Table 6. The clusters obtained by cutting dendrogram at an approximately height of 13.

| Cluster 1 (Low risk) | Cluster 2 (Medium risk) | Cluster 3 (High risk) |
|---|---|---|
| Agrigento | Ancona | Alessandria |
| Bari | Aosta | Avellino |
| Barletta-Andria-Trani | Arezzo | Bergamo |
| Benevento | Ascoli Piceno | Brescia |
| Brindisi | Asti | Como |
| Cagliari | Belluno | Cremona |
| Caltanissetta | Biella | Frosinone |
| Campobasso | Bologna | Lecco |
| Caserta | Bolzano | Lodi |
| Catania | Chieti | Mantua |
| Catanzaro | Cuneo | Milan |
| Cosenza | Fermo | Modena |
| Crotone | Ferrara | Monza and Brianza |
| Enna | Florence | Novara |
| Foggia | Forlì-Cesena | Padua |
| Isernia | Genoa | Parma |
| Latina | Gorizia | Pavia |
| Lecce | Grosseto | Pesaro and Urbino |
| Matera | Imperia | Piacenza |
| Messina | La Spezia | Reggio Emilia |
| Naples | L'Aquila | Rovigo |
| Nuoro | Livorno | Turin |
| Palermo | Lucca | Treviso |
| Potenza | Macerata | Venice |
| Ragusa | Massa-Carrara | Verona |
| Reggio Calabria | Oristano | Vicenza |
| Rome | Perugia | |
| Salerno | Pescara | |
| Sassari | Pisa | |
| Syracuse | Pistoia | |
| Taranto | Pordenone | |
| Trapani | Prato | |
| Vibo Valentia | Ravenna | |
| | Rieti | |
| | Savona | |
| | Siena | |
| | Sondrio | |
| | Sud Sardinia | |
| | Teramo | |
| | Terni | |
| | Trento | |
| | Trieste | |
| | Udine | |
| | Verbano-Cusio-Ossola | |
| | Vercelli | |
| | Viterbo | |